\documentstyle[12pt]{article}
\begin{document}
\def\la{\langle }
\def\ra{ \rangle }
\def\bea{\begin{eqnarray}}
\def\eea{\end{eqnarray}}
\def\beq{\begin{equation}}
\def\eeq{\end{equation}}
\def\gmmu{\gamma_{\mu}}
\def\gmf{\gamma _{5}}
\begin{center}

{\Large  Signatures of the Induced  $\Theta$ vacuum state   
in Heavy Ion Collisions}
\vskip1.0cm
      
{\Large Ariel Zhitnitsky}
\vskip0.5cm
{\it         Physics and Astronomy Department, 
        University of British Columbia, \\
Vancouver, BC V6T 1Z1, Canada } 
\vskip1.0cm
{\Large Abstract:\\}
\end{center}
\parbox[t]{\textwidth}{
It was   argued recently\cite{bfz}
that,  in general, an arbitrary induced $|\theta\ra^{ind}$-
vacuum state
would be created in the heavy ion collisions,
similar to the creation of the disoriented chiral condensate  
with an arbitrary isospin direction. It should be a large 
domain with a wrong $\theta^{ind} $ orientation
which will mimic the physics of the world when   the fundamental 
$\theta^{fund}\neq 0$. 
 I suggest a few simple observables   
which can (hopefully) be   measured on an event by event
basis at RHIC,
and which uniquely determine whether 
the induced $|\theta\ra^{ind}$ vacuum state is created.}

\vskip 0.4cm
\section{Introduction}

In a previous Letter\cite{bfz} it has been argued that a non-trivial 
$|\theta\ra^{ind}$ -vacua may be formed in   heavy ion collisions\footnote{
In what follows we  often omit the label ``ind" for the induced  $\theta$
vacuum state. We hope this state $|\theta\ra^{ind}$ will not be confused
with   $|\theta\ra^{fundamental}$ which is zero in our world and
which can not be changed in QCD. The simplest way to visualize
$|\theta\ra^{ind}$ is to assume that right after the QCD phase transition
the flavor singlet phase of the chiral condensate
is non-zero in a macroscopically large domain. This 
phase is identified with $\theta^{ind}$. This  identification
is a direct consequence of the transformation properties of
the fundamental QCD lagrangian under $U(1)_A$ rotations
by which the chiral singlet phase can be 
rotated away at the cost of the  appearence of the induced
$\theta^{ind}$ ( see  Sections 2.5 and 3.2 for detail 
explanations of the adopted terminology. See also \cite{TDLee}).}.
As is known, in the infinite volume limit
and in   thermal equilibrium the $|\theta\ra$ vacuum state
is the absolutely stable ground state  of a new world with 
  new physics  
quite  different from ours. 
In particular, P and CP symmetries are
strongly violated in this world. In spite of the fact that the $|\theta\ra$ vacuum
state has a higher energy (see below)
this state is stable 
due to the superselection rule:
There is no any gauge invariant observable  $\cal{A}$ in QCD, which would communicate between 
two different worlds, $\la\theta'|\cal{A}|\theta\ra\sim \delta(\theta-\theta')$.
Therefore, there are no   transitions between
these worlds\cite{gross}. The problem
of why $\theta=0$ in our world is known as the strong CP problem.  The best solution of this problem,
 which has passed the   test of time, is the introduction of the Peccei-Quinn symmetry\cite{PQ} and 
the corresponding pseudo-goldstone particle known as the  axion\cite{axion},
\cite{invisible}-\cite{dfsz},\cite{review}.  The axion solution of the strong 
CP problem suggests that the $\theta$ parameter of QCD is promoted to the 
dynamical axion field $\theta\rightarrow\frac{a(x)}{f_a}$ with very small coupling constant,  $1/f_a \ll 1$.
Once $\theta$ becomes the dynamical field, it 
automatically relaxes to $\theta=0$ as the
lowest energy state.

Of course, we do not expect that such a stable
$|\theta\ra$  state can be produced
in heavy ion collisions: However, we do expect
that locally, for a short period of time
due to the non-equilibrium dynamics after the
QCD phase transition, a large domain with wrong $|\theta\ra$ direction may be formed.  
This provides us with a unique opportunity to study
a new state of matter.
Numerical calculations of the evolution of the
chiral fields after the QCD phase transition
support this expectation\cite{bfz}.
An obvious question is how this 
$|\theta\ra$ vacuum state can be observed?
To be more concrete, the question
we are addressing in this letter is formulated in the following way:
``What kind of {\it  observable should be measured at RHIC }to confirm that
 a new ground state (and, therefore, a new set of excitations accompanying 
this ground state) has been created? ''.
The most natural answer 
to this question is that:    If  the axions 
 do exist, they  can be produced during the 
relaxation of  $|\theta\neq 0\ra^{ind}$
 to the lowest possible state with $\theta= 0$ as suggested in \cite{HZ},\cite{Melissinos}. 
 However,   the axion  production rate under the conditions which
can be achieved at RHIC
  will be too low in comparison with  the limit 
already achieved from the astrophysical and cosmological
 considerations\cite{axion}.
The second obvious choice for the observable which
can be a signal of   creation of the $\theta$ vacuum state
is  some CP odd correlation  as recently suggested in\cite{Pisarski}
\footnote{In the original papers\cite{Pisarski}, the authors 
  discussed not the induced $|\theta\ra^{ind}$ vacua which is the subject
of the present paper, but rather some metastable
(even in the infinite volume limit) vacua
which may (or may not) exist   within
the given QCD parameters  
$m_i,~ N_c,~ N_f, ~ \la 0|\bar{\Psi}_{i} \Psi_{i}|0 \ra $ as we know them.
It is not the  goal of this paper to comment on
whether   such metastable states are
likely /unlikely  to exist in Nature; or
likely /unlikely to be produced at RHIC. Rather we want
to emphasize that the CP odd observables suggested in
\cite{Pisarski} are sensitive to 
any $CP$ odd physics, not  necessary related to the 
metastable states introduced in\cite{Pisarski}.}.  
However, as Roberto Peccei noticed  recently\footnote{
I am thankful to Roberto for the discussions
on this subject,
which actually motivated the present search for an  alternative signal on 
the decay products in   the $|\theta\ra^{ind}$
 vacua background.} the signal could be considerably   
 washed out by   re-scattering of the pions and their interactions in 
the final states.
Indeed, as is well-known (see e.g. reviews\cite{Peccei}),
the final state interactions
will produce   complex phases which mimic true CP-odd effects.
In practice it is      quite difficult to overcome the difficulty
of separating a true CP violation from its simulation
due to the   final state interactions. For example,
it is well known that the final state interactions
is the most difficult obstacle in the study of   
CP violation
in $K^{0}(\bar{K^{0}})\rightarrow \pi^{\pm}\mu^{\mp}\nu$ decays by measuring the muon 
transverse polarization $ \xi_i$ (defined  by the correlation
$(\vec{p_{\pi}} \times\vec{p_{\mu}}\cdot \vec{\xi})$). Fortunately, the final state 
interactions in the charged kaon decays $K^{+}({K^{-}})\rightarrow \pi^{0}\mu^{\pm}\nu$ 
are strongly suppressed\cite{arz}, and therefore a measurement of this correlation in 
the ongoing experiment at KEK\cite{KEK} on the level $>10^{-6}$ would imply a new source 
of CP violation.
In general, one should expect that similar mimicry of true CP violation due to the 
final state interactions also occurs in the heavy ion collisions.
The main goal of this letter is to offer
some new observable which:\\ a)  can be measured at RHIC;\\
b)	will be the signal of the new physics of the $\theta$ vacua; \\
c)	does not suffer from the deficiencies mentioned above.\\
Before I explain the main idea of the construction of  such an observable, I would 
like to explain the low-energy physics in the $\theta$ world.

\section{Low Energy Physics of  the $ \theta $ World}
In this section (except subsection 2.5) 
we study the phenomenology of the world when $\theta^{fund}\neq 0$.
It will give us some useful hints about CP odd physics in the unusual environment with $\theta\neq 0$.
 
The starting point of our analysis is the low energy effective lagrangian which 
reproduces the anomalous conformal and chiral Ward Identities.
The corresponding construction in the large $ N_c$ limit has been known 
for a long time\cite{Wit2},\cite{WV}. The generalization of the construction
of ref.\cite{Wit2},\cite{WV} for finite $N_c$ was given  in \cite{QCD},\cite{QCDanalysis},
and we shall use formulae from the papers
\cite{QCD},\cite{QCDanalysis}.  However, we should remark
at the very beginning   that all 
local properties of the effective 
lagrangians for finite and
infinite $N_c$    are very much the same. Small quantitative differences in 
local physics between the description of \cite{Wit2},\cite{WV} on the one hand and 
the description of\cite{QCD},\cite{QCDanalysis} on the other hand, do  not alter the qualitative 
results which follow.
First  I discuss the  properties 
of the unusual $\theta$ vacua itself. Discussions of the  properties of 
the Goldstone bosons in the $\theta$
vacuum  background will follow after that. 

\subsection{  Effective chiral  lagrangian and  properties
of $\theta$ vacua}
In the effective Lagrangian approach, the light matter fields are 
described by the unitary matrix $U_{ij}$ corresponding to the phases
of the chiral condensate $
\la 0|\bar{\Psi}^{i}_L \Psi^{j}_R|0 \ra =
|\la 0|\bar{\Psi}_L \Psi_R|0 \ra |U_{ij}$.
In terms of $U$ the effective potential
for arbitrary $\theta$ takes the form
\cite{QCD},\cite{QCDanalysis}:
\beq
\label{9}
V( U, \theta) = - E \cos \left[  \frac{1}{N_c} ( \theta -
i \log Det \, U ) \right] - \frac{1}{2} \,  
Tr \, (MU + M^{+}U^{+} ) \; ,
\eeq
where $M$ is the diagonal mass matrix for quark masses defined with their condensates
($M_i=-  diag (m_i\la \bar{\Psi}_{i} \Psi_{i} \ra )$ ) and 
$  E =  \la b \alpha_s /(32 \pi) G^2 \ra $, with $ b=11N_c/3-2N_f/3$,
is the vacuum energy (``cosmological" term ) required by the conformal 
anomaly. Expanding the cosine (this corresponds to the expansion
in $   1/N_c $) we recover
   the result of \cite{Wit2} at lowest order 
in $ 1/N_c $  together with 
the constant term $E$:
\beq
\label{VVW}
V( U, \theta, N_c\rightarrow\infty) = = - E - \frac{ \la \nu^2 \ra_{YM} }{2}
( \theta - i \log Det \, U )^2 - \frac{1}{2} \, Tr \, (MU + M^{+}U^{+} ) + \ldots \; , 
\eeq 
where we used the fact that in the  large 
$ N_c $ limit $  E(\frac{1}{N_c})^2 = - \la \nu^2 \ra_{YM} $ where $ \la \nu^2 \ra_{YM}< 0 $ 
is the topological susceptibility in pure YM theory. 
Corrections in $ 1/N_c $ stemming from Eq.(\ref{9}) constitute a new 
result of ref.\cite{QCD}.
 First of all,  let me  review   the 
picture of the vacuum structure   for  $ \theta \neq 0$
stemming  from the effective potential (\ref{9}).    In the next 
subsection I review some properties of the  Goldstone modes in  the $ \theta \neq 0$ 
background.
To study the vacuum properties
it is convenient to   parametrize the 
fields $U$ as $ U= diag( \exp i \phi_q ) $ 
such that the potential (\ref{9}) takes the form:
\beq
\label{13}
V = - E \cos \left( \frac{1}{N_c} \theta -
\frac{1}{N_c} \sum \phi_{i}  \right) 
- \sum M_{i} \cos \phi_i  \; \; , \; \;    
\eeq
The minimum of this potential is determined by the following equation:
\beq
\label{15} 
\frac{1}{N_c}\sin \left( \frac{1}{N_c} \theta - \frac{1}{N_c} 
\sum \phi_i \right) =  \frac{M_i}{E} \, \sin \phi_i \; \; , \; \; i = 1,
\ldots, N_f .
\eeq 
At lowest order in
$ 1/N_c $ this equation coincides with that of \cite{Wit2}. For general 
values of $ M_{i} / E $, it is not possible to solve Eq.(\ref{15}) 
analytically.
However, in the realistic case $ \varepsilon_{u},
 \varepsilon_{d} \ll 1 \, , \, \varepsilon_s \sim 1 $ where 
$ \varepsilon_{i} = \frac{N_cM_{i}}{E }$, the approximate solution 
can be found: 
\bea
\label{21}
\sin \phi_{u} &=& 
\frac{ m_d \sin \theta }{ [m_{u}^2 + m_{d}^2 +
2 m_{u} m_{d} \cos \theta ]^{1/2} } + O(\varepsilon_{u},
\varepsilon_{d}) \; , \nonumber \\
\sin \phi_{d} &=& 
\frac{ m_u \sin \theta }{ [m_{u}^2 + m_{d}^2 + 2 m_{u} m_{d}
 \cos \theta ]^{1/2} } + O(\varepsilon_{u}, \varepsilon_{d}) \; ,
  \\ \sin \phi_{s}&=&  O(\varepsilon_{u},\varepsilon_{d})  \nonumber \; .
\eea
This solution   coincides with the 
one of Ref.\cite{Wit2} to   leading order in 
$ \varepsilon_{u},\varepsilon_{d} $. 
In what follows for the numerical 
estimates and for simplicity 
we shall use    the $SU(2)$ limit $m_u=m_d \neq m_s$ where the solution 
(\ref{21})  can be approximated as:
\bea
\label{vacuum}
\phi_{u} \simeq  \frac{\theta}{2} ,~~~
\phi_{d} \simeq  \frac{\theta}{2} ,~~~
\phi_{s} \simeq 0 ,~~~
0\leq	\theta < \pi \nonumber\\
\phi_{u} \simeq \frac{\theta+2\pi}{2} ,~~~
\phi_{d}  \simeq \frac{\theta-2\pi}{2} ,~~~
\phi_{s} \simeq 0 , ~~~ 
\pi\leq	\theta < 2\pi .
\eea

Once solution (\ref{vacuum}) is known, 
one can calculate the vacuum energy
and topological charge density  
$Q=\langle 0 | \frac{\alpha_{s}}{8 \pi } G \tilde{G}
|0 \rangle$
as a function of $\theta$. In the limit
$m_u=m_d\equiv m,~~
\la \bar{d} d \ra = \la \bar{u} u \ra \equiv \la \bar{q} q \ra $ one has:
\bea
\label{energy}
V_{vac}(\theta)\simeq V_{vac}(\theta=0)-2m|\la \bar{q} q \ra |
(|\cos\frac{\theta}{2}|-1) \nonumber \\
\langle \theta | \frac{\alpha_{s}}{8 \pi } G \tilde{G}
| \theta \rangle =-\frac{\partial V_{vac}(\theta)}{\partial\theta}=
-m|\la 0| \bar{q} q |0 \ra |\sin\frac{\theta}{2},
\eea
As expected, the $\theta$ dependence appears only in combination
 with $m$ and goes away in the chiral limit.
One can also calculate the chiral condensate
$ \la \bar{\Psi}^{i}_L \Psi^{i}_R \ra $ in the $\theta$ vacua 
using   solution (\ref{vacuum}) for vacuum phases:
\beq
\label{chiral}
\la \theta |
\bar{q} q |\theta \ra 
=\cos\frac{\theta}{2} \la 0|\bar{q} q |0\ra_{\theta=0},
~~~~
\la \theta |\bar{q}i\gamma_5 q |\theta\ra 
=-\sin\frac{\theta}{2} \la 0|\bar{q} q |0 \ra_{\theta=0}
\eeq	
Both expressions (\ref{energy}) and  (\ref{chiral})
explicitly demonstrate that P, CP symmetries 
are explicitly broken in the $\theta$ vacuum background. This  has very important 
phenomenological consequences which will be discussed in the next section.
The crucial point for the analysis  presented above
is the fact  that the $\theta$ parameter appears
in the effective lagrangian (\ref{9}) only in 
the combination,
$( \theta -
i \log Det \, U )$.  This is a direct consequence 
of the transformation properties of the chiral
fields under $U(1)_A$ rotations. 
To convince the reader that (\ref{9}) does indeed represent the
anomalous effective low energy
Lagrangian, three of its most salient features are
listed below :\\[0mm]
\indent i)
Eq. (\ref{9}) correctly reproduces the Witten-
Di Vecchia -Veneziano effective
chiral Lagrangian
\cite{Wit2} in the large $ N_c $ limit; \\[0mm]
\indent ii)
it reproduces the anomalous conformal and chiral Ward identities of $QCD$;\\[0mm] \indent iii) 
it reproduces the known dependence in $\theta$  \cite{Wit2} up to small corrections.
Accordingly, it leads to the correct $2\pi$
periodicity of observables.  

\subsection{Goldstone bosons in $\theta$ world  }
 In this subsection I would like to describe
 some unusual properties of the Goldstone modes in the unusual $\theta$ environment.
I  refer the interested  readers to the original papers \cite{QCD},\cite{QCDanalysis} for
 a detailed  description of the approach.  Here I present some old/new    results which can be 
easily derived from the general expressions given in ref.\cite{QCDanalysis}. 
To study the properties of the pseudo-goldstone 
bosons, we parametrize the chiral matrix (\ref{9}) in the form
\beq
\label{20k}
U = U_{0} \, \exp \left[ i \sqrt{2} \, \frac{\pi^{a} 
\lambda^{a} }{f_{\pi}}  + 
i \frac{ 2}{ \sqrt{N_{f}} } \frac{ \eta'}{ f_{\pi}}  \right] 
\; ,
\eeq where $\lambda^a$ are Gell-Mann matrices of $SU(N_f=3)~$, $\pi^a$ is the octet of 
pseudo-goldstone bosons, and $ f_{\eta'}\simeq f_{\pi}\simeq 133 MeV$.
In this formula $ U_0 $ solves the minimization equations for the effective potential 
(\ref{9}), and the fields $ \pi^{a} \, , \, \eta' $ all have vanishing vacuum expectation values.
In what follows we shall use the approximate solution
(\ref{vacuum}) for the vacuum phases
$\phi_i$. A simple calculation
of the mass matrix  in the $SU(2)$ limit when  $m_u=m_d\equiv m$ 
yields the following result   for an arbitrary value of 
$ \theta $ (for the general case $m_u\neq m_d$ see\cite{QCDanalysis}):
\bea
\label{20g}
m_{\pi}^2 = \frac{4}{f_{\pi}^2} ( M_q|\cos \frac{\theta}{2}| ),
~~~~~~~m_{\eta}^2 = \frac{4}{3 f_{\pi}^2} (M_q|\cos \frac{\theta}{2}|
+ 2 M_s  )  \nonumber \\
m_{\eta'}^2 = 4  \frac{E N_f}{f_{\pi}^2 N_c^2} + \frac{4}{N_f} 
\frac{1}{f_{\pi}^2}
\left( 2M_q|\cos \frac{\theta}{2}|+M_s \right), \\
m_{\eta,\eta'}^2 = m_{\eta' ,\eta}^2  =  2 
\sqrt{ \frac{2}{3 N_{f}} }\frac{1}{f_{\pi}^2}
\left( 2M_q |\cos \frac{\theta}{2}| -2M_s )\right) \nonumber   
\eea
In this isospin  limit  $m_u=m_d\equiv m$,  the $ \pi^{0}$
does not mix with $ \eta $ or $  \eta' $,  and  $m_{\pi^0} $
coincides with the physical mass of the $\pi^0 $.    At the same time 
$ \eta $ and $  \eta' $  do mix  with each other.  The mixing  is determined by the following    matrix: 
\bea
\label{300}
m_{\eta-\eta'}^2 = \left( \begin{array}{cl} 
- \frac{4}{3 f_{\pi}^2}( 2 m_s \la \bar{s}
s \ra + m \la \bar{q} q \ra|\cos \frac{\theta}{2}|  ) &  
\frac{4 \sqrt{2}}{ 3 f_{\pi}^2 }
(  m_s \la \bar{s}
s \ra - m \la \bar{q} q \ra |\cos \frac{\theta}{2}|) \\
\frac{4 \sqrt{2}}{ 3 f_{\pi}^2}
(  m_s \la \bar{s}
s \ra - m \la \bar{q} q \ra |\cos \frac{\theta}{2}|) &  - \frac{4}{3 f_{\pi}^2} 
(  m_s \la \bar{s}
s \ra +2 m \la \bar{q} q \ra |\cos \frac{\theta}{2}|) + 4  \, 
\frac{E N_f}{f_{\pi}^2 N_c^2}
\end{array} \right)
\eea
which coincides  with an accuracy $ O(m^2) $ with the matrix given
by Veneziano at $\theta=0$\cite{WV}
\bea
\label{301}
m_{\eta-\eta'}^2 = \left( \begin{array}{cl} 
\frac{1}{3} ( 4 m_{K}^2 - m_{\pi}^2 ) 
&  -  \frac{2 \sqrt{2}}{ 3 }  (  m_{K}^2 - m_{\pi}^2 ) \\ -  
\frac{2 \sqrt{2}}{ 3 }  (  m_{K}^2 - m_{\pi}^2 ) & \frac{2}{3} m_{K}^2 +
 \frac{1}{3} m_{\pi}^2 + \frac{ \chi}{N_c} \end{array} \right).
\eea
The only  difference between eq.(\ref{300})
and Eq.(\ref{301}) is that the topological
susceptibility $ \sim \chi $ in pure YM theory in the latter 
is substituted by the term proportional to the gluon condensate
in   real QCD in the former.  

A few remarks are in order.
First, from Eq.(\ref{20g}) one could naively think that
the $\pi$ meson   becomes massless at $\theta=\pi$.
This  is a consequence of our approximation $m_u=m_d$.
For $m_u\neq m_d$ the $\pi$ meson never becomes massless,
and  in the vicinity $\theta\simeq \pi$
one should use the exact formula $m_{\pi}^2=
\frac{2}{f_{\pi}^2}(M_u\cos(\phi_u)+ M_d\cos(\phi_d))$ with exact solutions 
(\ref{21}) for phases $\phi_u , ~\phi_d$ ( see\cite{QCDanalysis}). However, 
the  observation that the masses of the pseudo-goldstone bosons get 
corrections of  the order of  $1$  in  the $\theta\neq 0$ background   is 
absolutely correct.
  In particular, the  $\pi$ meson could   easily  have a mass of, let us say,  
$95 ~MeV$ for a certain value of $\theta$ rather than $135~ MeV$ from 
the Particle Data Booklet. Masses of all other hadrons are also 
influenced by the $\theta$ vacua.  However,  this influence is not 
as profound as for pseudo-goldstone bosons because all effects 
due to the $\theta\neq 0$ background are proportional to $m$ 
and must vanish in the chiral limit.  For pseuodo-Goldstone bosons 
and  the $\eta'$ meson this effect can be calculated exactly in the  
 limit $m\rightarrow 0$: 
\bea
\label{mass}
\frac{m_{\eta'}^2(\theta=0)-
m_{\eta'}^2(\theta\neq 0)}{m_{\eta'}^2(\theta=0)}=
\frac{8}{3 f_{\pi}^2m_{\eta'}^2} 
m |\la \bar{q} q \ra |\left(1-|\cos \frac{\theta}{2}|\right)>0
\nonumber \\
\frac{m_{\eta}^2(\theta=0)-
m_{\eta}^2(\theta\neq 0)}{m_{\eta}^2(\theta=0)}=
\frac{4}{3 f_{\pi}^2m_{\eta}^2} 
m |\la \bar{q} q \ra |\left(1-|\cos \frac{\theta}{2}|\right)>0 \\
\frac{m_{K}^2(\theta=0)-
m_{K}^2(\theta\neq 0)}{m_{K}^2(\theta=0)}=
\frac{2}{ f_{\pi}^2m_{K}^2} 
m |\la \bar{q} q \ra |\left(1-|\cos \frac{\theta}{2}|\right)>0
\nonumber
\eea
For all other heavy hadrons one can not present analogous calculations;
however,  one should expect a similar,
quite substantial effect 
$$   \frac{m \la \bar{q} q \ra }{f_{\pi}^2}\sim (100 ~MeV)^2$$ for all hadrons. 

My first remark was about masses (spectrum), see discussions above; my second remark
 is about quantum numbers.
Namely, I would 
like to argue that all hadrons  in the 
$ \theta $-vacuum environment cease to 
carry  definite P, CP parities
but  instead become mixtures of the   states 
with different quantum numbers.
This remark becomes  quite obvious if
one remembers that the $\theta$ vacuum 
ground state with the chiral and gluon condensates given by formulae (\ref{chiral}) and (\ref{energy}) 
correspondingly, is not invariant under these symmetries. 
Therefore, in general  one should expect that
all excitations in this background are 
not eigenstates of P and CP parities. 
Therefore:
in the presence of a non-zero angle $ \theta $, the pseudo-goldstone bosons cease to be the pure 
pseudoscalars  and acquire scalar components.
This    mixing of states
with different parities
has been known since\cite{svz}
where   it was explicitly demonstrated that the charmonium levels with quantum numbers
 $0^+$ and $0^-$ do mix.  In this case the calculations are under theoretical control 
(due to the large mass $m_c$ of the charmed quark), and the result for the mixing is
 expressed in terms of  the CP-odd gluon condensate 
$ \langle \theta | \frac{\alpha_{s}}{4 \pi } G \tilde{G} | \theta \rangle =
 -2m|\la 0| \bar{q} q |0 \ra |\sin\frac{\theta}{2}$, see (\ref{energy}).
It was also demonstrated in ref.\cite{svz} that the 
$\eta\rightarrow\pi\pi$ decay is allowed  
in the $\theta$ vacuum background.
This decay   also can be interpreted
as a result of mixing pseudo-Goldstone bosons $0^-$ with  a scalar particle $0^+$ 
which can easily decay to two $\pi$ mesons.
Due to the 
phenomenological importance of this result
as a definite  signature of the produced $|\theta\ra^{ind}$   state in heavy ion collisions, 
we derive the decay rate for $\Gamma(\eta\rightarrow\pi\pi)$ in a  separate
subsection below.
We express the rate
$\Gamma(\eta\rightarrow\pi\pi)$ exclusively
in terms of the chiral vacuum condensate (\ref{chiral}).
At small $\theta$ our formula coincides with the
one presented long ago in ref.\cite{svz}.

\subsection{$\eta\rightarrow\pi\pi$ decay}
As is known, this decay is strongly
suppressed in our world with
$\theta=0$. Indeed, CP parity for the initial state is $(-)$, while for the final 
state it is $(+)$.
This is the main reason for  a very small full width
$\Gamma\simeq 1.2 KeV$ for $\eta$ meson
which can decay strongly only to  $ 3\pi$ mesons with a strong isospin 
suppression (three $\pi$ mesons can not be in the $I=0$ state) as well as  a phase volume suppression.
The starting point of our calculations  is the 
following matrix element:
\beq
\label{eta}
A(\eta\rightarrow\pi\pi)=\la 2\pi|-m_u\bar{u}u-m_d\bar{d}d|\eta\ra  .
\eeq
 Our goal is to estimate this matrix element in the limit of vanishing 
$4$-momenta of the $\pi$ and $\eta$ mesons.  Therefore, one can use the 
standard PCAC-technique.
The result is:
\beq
A(\eta\rightarrow\pi\pi)=\la 2\pi|-m_u\bar{u}u-m_d\bar{d}d|\eta\ra \simeq 
\frac{2}{\sqrt{6}}\frac{(m_u+m_d)
\la \theta | \bar{u}i\gamma_5 u+\bar{d}i\gamma_5 d |\theta\ra }
{f_{\pi}^3}\simeq \nonumber 
\eeq
\beq
\label{eta1}
-\frac{8}{\sqrt{6}}\frac{m
\sin\frac{\theta}{2} \la 0|\bar{q} q |0 \ra_{\theta=0}}{f_{\pi}^3}, 
\eeq
 where at the last step we used
 formula  (\ref{chiral})
for  the CP odd chiral condensate which is not zero
in the $\theta$ world.
In this derivation
we neglect all corrections proportional
to $m_u-m_d$ having in mind that at the very end we shall use the approximate solution (\ref{vacuum}) 
for the chiral phases and masses (\ref{mass}) rather than the exact expression (\ref{21}).
Now we are ready to estimate the decay rate:
\beq
\Gamma(\eta\rightarrow\pi\pi)=\frac{|A(\eta\rightarrow\pi\pi)|^2}{16 \pi
m_{\eta}(\theta)}\left(1-\frac{4m_{\pi}^2(\theta)}{m_{\eta}^2(\theta)}
\right)^{\frac{1}{2}}\simeq \nonumber
\eeq
\beq
\label{eta2}
\simeq \frac{2}{3\pi m_{\eta}(\theta)}\left(\frac{m
\sin\frac{\theta}{2} \la 0|\bar{q} q |0 \ra}{f_{\pi}^3}\right)^2
\left(1-\frac{4m_{\pi}^2(\theta)}{m_{\eta}^2(\theta)}
\right)^{\frac{1}{2}} ,
\eeq
with all masses to be calculated in the $\theta$ vacua (\ref{mass}). 
Formula (\ref{eta2}) reduces to the corresponding expression 
of ref.\cite{svz} in the limit $\theta\rightarrow 0$
when the standard PCAC relation
$m | \la 0 |\bar{q} q |0 \ra |\rightarrow \frac{1}{4}
m_{\pi}^2f_{\pi}^2$ is substituted 
into eq.(\ref{eta2}).
For numerical estimates we neglect the
$\theta$ dependence of masses and arrive
at the following decay rate:
\beq
\label{rate}
\Gamma(\eta\rightarrow\pi\pi)\sim 0.5 MeV\cdot
(\sin\frac{\theta}{2})^2 ,
\eeq
which essentially determines the full width
of the $\eta$ meson in the $\theta\neq 0$ world.

\subsection{$\eta '\rightarrow\pi\pi$ decay}
One can carry out  similar calculations for  the $\eta'\rightarrow\pi\pi$ decay as well.
This decay is also strongly suppressed  for  the  same  reasons   discussed above for 
the $\eta\rightarrow\pi\pi$ decay.
The full width of $\eta '$ meson (from particle data booklet),
$\Gamma(\eta ')\simeq 0.2 MeV$,  
is   very  small in comparison with
the ``normal" width $(\sim 100 MeV)$ which  one
could expect for a strongly interacting particle
with mass $\sim 1 GeV$. This suppression is 
 due to the  small phase volume  
of the only allowed strong decay,  $\eta'\rightarrow\eta\pi\pi$.

The starting point of our calculations of
$\Gamma(\eta '\rightarrow\pi\pi)$   
with  $\theta\neq 0$ is   the 
following matrix element
\beq
\label{eta3}
A(\eta '\rightarrow\pi\pi)=\la 2\pi|-m_u\bar{u}u-m_d\bar{d}d|\eta'\ra .
\eeq
Our goal is to estimate this matrix element in the limit
of vanishing $4$-momenta of the $\pi$ mesons.
We use the standard PCAC-technique as we did above
for $\eta\rightarrow\pi\pi$ decay.
The result of  this calculation can be presented in
terms of    the following
$\eta '$ matrix element:
\beq
\label{eta4}
A(\eta '\rightarrow\pi\pi)=\la 2\pi|-m_u\bar{u}u-m_d\bar{d}d|\eta '\ra \simeq  \frac{(m_u+m_d)
\la 0 | \bar{u}  u+\bar{d}  d |\eta '\ra_{\theta} }{f_{\pi}^2}  .
\eeq
In our world with $\theta=0$ this matrix element 
is obviously zero. However, as   discussed above, 
due to the mixing of states with different parities
in the $\theta\neq 0$ background  this matrix element is
expected to be nonzero
in  the $\theta\neq 0$ vacuum state\footnote{
it must vanish in $\theta\rightarrow 0$
limit though. }. 

The simplest way to evaluate the $\eta'$ matrix element (\ref{eta4}) is to make use of 
our knowledge of the  vacuum condensates (\ref{energy}, \ref{chiral})in the $\theta\neq 0$ world.
One can differentiate $\langle \theta | \frac{\alpha_{s}}{8 \pi } G \tilde{G} | \theta \rangle $ 
with respect to $m$ in order to derive the following low energy theorem which will be used for the 
estimation of the matrix element (\ref{eta4}) in what follows.
\bea
\label{theorem}
(\frac{\partial}{\partial m_u}+\frac{\partial}{\partial m_d}) \langle \theta | 
\frac{\alpha_{s}}{8 \pi } G \tilde{G} | \theta \rangle =-i~ \lim_{q\rightarrow 0}
\int dx e^{iqx}\la \theta | (\bar{u} u(x) +\bar{d} d(x)) , \frac{\alpha_{s}}{8 \pi } G \tilde{G}(0)
| \theta\ra  =  \nonumber \\ 
=\frac{1}{2}
\la 0| \bar{u} u + \bar{d} d  |0 \ra \sin\frac{\theta}{2},
~~~~~~~~~~~~~~~~~~~~~~~~~~~~~~~~~
\eea
where at the last step we used our knowledge 
of the  $\theta$ dependence  for the vacuum condensate
$\langle \theta | \frac{\alpha_{s}}{8 \pi } G \tilde{G}
| \theta \rangle  =
m\la 0| \bar{q} q |0 \ra \sin\frac{\theta}{2}$, see eq.
(\ref{energy}).
The same relation can be obtained 
in a different way, by differentiating
$\la \theta |
\bar{q} q |\theta \ra $ with respect to $\theta$. Indeed,
\bea
\label{theorem1}
-\frac{\partial}{\partial \theta}\la \theta|
\bar{u} u + \bar{d} d  |\theta\ra = 
i~ \lim_{q\rightarrow 0}\int dx e^{iqx}\la \theta | (\bar{u} u(0) +\bar{d} d(0)) , \frac{\alpha_{s}}{8 \pi } G \tilde{G}(x)
| \theta\ra  =   \nonumber\\
=\frac{1}{2}
\la 0| \bar{u} u + \bar{d} d  |0 \ra \sin\frac{\theta}{2},
~~~~~~~~~~~~~~~~~~~~~~~~~~~~~
\eea
where at the last step we used our knowledge 
of the  $\theta$ dependence  for the vacuum condensate
$\la \theta |
\bar{q} q |\theta \ra 
=\cos\frac{\theta}{2} \la 0|\bar{q} q |0\ra$, see eq. (\ref{chiral}).  From 
this low-energy theorem one can easily estimate the matrix element 
(\ref{eta4}) we are interested in.
Indeed, using the standard dispersion relations and 
keeping in the
imaginary part the contribution of the $\eta'$ meson only
( justified in the large $N_c$ limit),
we arrive at the following estimate
for the $\eta'$ matrix element:
\beq
\label{eta5}
\langle 0 | \frac{\alpha_{s}}{8 \pi } G \tilde{G}
| \eta' \rangle_{\theta}
\cdot \la \eta' | \bar{u}  u+\bar{d}  d |0\ra_{\theta}=
m_{\eta'}^2(\theta)\frac{1}{2}
\la 0| \bar{u} u + \bar{d} d  |0 \ra \sin\frac{\theta}{2}.
\eeq
In the chiral limit one can neglect
the $\theta$- dependence in the mass term 
$m_{\eta'}^2 (\theta)\simeq m_{\eta'}^2 (\theta=0) +0(m) $   and 
in the matrix element 
$\langle 0 | \frac{\alpha_{s}}{8 \pi } G \tilde{G}
| \eta' \rangle_{\theta}\simeq 
\langle 0 | \frac{\alpha_{s}}{8 \pi } G \tilde{G}
| \eta' \rangle_{\theta=0} +0(m)  $. Therefore, 
for estimation purposes,  one can 
use the corresponding values calculated
for the $\theta=0$ vacuum. 
\beq
\label{eta6}
\la  0 | \bar{u}  u+\bar{d}  d |\eta '\ra_{\theta} =
-\frac{\sqrt{3}}{f_{\pi}(0.5-0.8)}\cdot
\la 0| \bar{u} u + \bar{d} d  |0 \ra \sin\frac{\theta}{2}
\eeq 
where we used the estimate of  the  matrix element
$\langle 0 | \frac{\alpha_{s}}{4 \pi } G \tilde{G}
| \eta' \rangle \simeq \frac{(0.5-0.8)}{\sqrt{3}}f_{\pi}m_{\eta'}^2 $ from 
ref.\cite{svz2}. Substituting this expression into (\ref{eta4}), we arrive at 
the following estimate for the amplitude  of  $\eta'\rightarrow\pi\pi$:
\beq
A(\eta'\rightarrow\pi\pi)\simeq
-\frac{4 \sqrt{3}}{(0.5-0.8)}\frac{m
\sin\frac{\theta}{2} \la 0|\bar{q} q |0 \ra_{\theta=0}}{f_{\pi}^3}, \eeq which 
is the same order of magnitude as $A(\eta\rightarrow\pi\pi)$ (\ref{eta1}).
This value
for the amplitude translates to the following  
estimate for the decay rate
\beq\label{eta7}
\Gamma(\eta'\rightarrow\pi\pi)\simeq
\frac{|A(\eta'\rightarrow\pi\pi)|^2}{16 \pi
m_{\eta'}} 
\sim \frac{6}{\pi m_{\eta'}}\left(\frac{m
\sin\frac{\theta}{2} \la 0|\bar{q} q |0 \ra}{f_{\pi}^3}\right)^2
\sim 2~MeV (\sin\frac{\theta}{2})^2,
\eeq
which is almost an order of magnitude larger
than the full width of the $\eta'$ meson in the 
$\theta=0$ world. What is more important, 
this width is exclusively due to the 
CP odd decay $\Gamma(\eta'\rightarrow\pi\pi)$.

\subsection{Forming of the induced $\theta $ vacua at RHIC}
Now we want to argue\cite{bfz} that the induced $|\theta\ra^{ind}$
vacuum state with an effective $\theta^{ind}\neq 0$  ( which was treated so far
as a pure theoretical construction) can, nevertheless, be 
experimentally produced and studied 
in the real world in the relativistic heavy 
ion collider.
Of course, we do not expect that  a stable
$|\theta\ra^{ind}$  state can be produced
in heavy ion collisions; however, we do expect
that a relatively stable and 
sufficiently large
    domain with a wrong
induced $|\theta\ra^{ind}$ orientation may be formed due to the non-equilibrium 
dynamics after the QCD phase transition. If this  is indeed the case, such a study 
gives us a unique opportunity  to mimic the physics of the world when   the 
fundamental $\theta^{fund}\neq 0$. This situation  is believed to  have  occurred 
during the QCD phase transition in the development of the early Universe.

The idea is very similar to the old idea of creation of the Disoriented Chiral 
Condensate (DCC) in heavy ion collisions \cite{DCC} ,\cite{RW}, \cite{Rajagopal}
 (see also nice review \cite{Boyanovsky} for a discussion of DCC as an example 
of an out of equilibrium phase transition).
DCC refers to regions of space (interior) in which the chiral condensate points 
in a different direction from that of the ground state (exterior), and separated 
from the latter by a hot shell of debris. 
Our starting point is the conjecture
that a classically large domain with a nonzero
$U(1)_A$ chiral phase may be formed in a  heavy ion collisions, i.e. we assume
 that the expectation value for the chiral condensate in a sufficiently large 
domain has, in general,  a nonzero $U(1)$ phase:
$\la \bar{\Psi}_L \Psi_R \ra \sim
e^{i\phi^{singlet}}| \la \bar{\Psi} \Psi \ra | $.
This  
phase is identified with $\theta^{ind}$. Such an identification is a direct 
consequence of the transformation properties of the fundamental QCD lagrangian 
under $U(1)_A$ rotations when the chiral singlet phase can be rotated away at the 
cost of the introduction  of the induced  $\theta$, $\theta^{ind}=N_f\phi^{singlet}$.
The production of non-trivial $|\theta\ra^{ind}$-vacua would occur in
much the same 
way as discussed above for DCC.
The new element is that in addition to 
chiral fields differing from their true vacuum values the induced $\theta$-parameter,
 which is zero in the real world, becomes effectively nonvanishing in the
 macroscopically large domain.
Of course, 
once such a conjecture is made
( the classically large domain with a nonzero
$U(1)_A$ chiral phase is formed) one has to check
that this assumption  is   
  self-consistent. Namely, one should check
that (initially)   randomly distributed phases $\phi_i$
in the $\theta^{ind}$-induced background
will, indeed, relax to the same    non-zero value 
$\phi_i\rightarrow\sim \frac{\theta_{ind}}{N_f}$ in a sufficiently short period of time.
The numerical study of ref.\cite{bfz}
suggests this  is   indeed the  case.

We want to pause here to explain
the terminology we adopt regarding 
the induced  $|\theta\ra^{ind}$ vacua.
Of course the fundamental parameter $\theta$  which enters the fundamental 
QCD lagrangian is zero in our world.  One should remember, however, that 
the fundamental $\theta=0$ is a combination of two  pieces: the first part 
is related to the  original term $\theta Q= \theta\frac{\alpha_{s}}{8 \pi }
 G^a \tilde{G}^a$; the second part is related to the mass matrix $\cal M$ which, 
in general, may have an arbitrary phase.
The $U(1)_A$ rotation of the $\Psi$ fields
brings this matrix to
a canonical form, producing however an additional contribution to the 
$\theta$ term\footnote{By the way, the famous strong CP problem is
 formulated in these terms as follows:
Why do these two contributions cancel each other
with precision better than $10^{-9}$?}.
Our induced  $\theta^{ind}$   can be considered as the second term related to the 
rotation of the $\Psi$ fields.
After this point we can apply
the same philosophy as for DCC.  
The chiral fields
$ \phi_i $ are allowed to take random values as discussed above, and 
they begin to roll toward the true solution $\bar{\phi_u}
\simeq \bar{\phi_d}\simeq\theta/2, \bar{\phi_s}\simeq 0$ (\ref{vacuum}) 
and of course overshoot  it. The situation is very similar to what 
was described for the DCC with the only difference that  in general 
we expect an  arbitrary $|\theta\ra^{ind}$-disoriented 
state to be created 
in heavy ion collisions,
not necessarily the $|\theta=0\ra$ state.  
It is assumed that the
rapid expansion of the high energy shell leaves behind an effectively zero 
temperature region in the interior which is isolated from the true  vacuum.
The high temperature non-equilibrium evolution is very suddenly stopped, 
or ``quenched'', leaving the interior region in a non-equilibrium initial 
state that then begins to evolve according to (almost) zero temperature 
 Lagrangian  dynamics.
Therefore starting  from an {\it initial non-equilibrium}
state we can study the 
behavior of the chiral fields using the 
{\it zero temperature} equations of motion.
These equations   are non-linear and cannot be solved analytically
but the  numerical analysis can be done  and has
been presented in ref.\cite{bfz}, where 
the equations of motion for the phases of the chiral condensate 
in the theory with two
quark flavors have been studied:
\beq
\label{equation}
\ddot{\phi}_i - \nabla^2 {\phi_i} + \gamma \dot{\phi_i} 
+ \frac{d}{d\phi_i} 
V(\phi_j,\theta)=0 \; \; \; \; \; i=u,d.
\eeq
Here $\nabla^2$ is a three dimensional spatial derivative and the
 potential is given in (\ref{13}).  
Emission of pions and expansion of the domain will contribute to 
the damping, $\gamma$, as might other processes.  The contribution of  these unknown 
effects was simulated in ref.\cite{bfz} by including a damping term 
with a reasonable value for 
the damping constant,  $\gamma \sim\Lambda_{QCD}\sim 200 ~ MeV$.
The results of this numerical 
study can be summarized  in the following way:

a)The $|\theta\ra^{ind}$ domain does  form.
The production of a non-trivial $\theta$  is indicated by the 
fact that  
the chiral fields relax to constant non-zero values on a time scale
over which spatial oscillations of the fields vanish. In other words,  
the zero mode   settles down to a non-zero
value $\simeq \frac{\theta}{2}$.
At the same time,   
all higher momentum modes vanish extremely rapidly and are negligible
 long before the zero mode settles down to its equilibrium value.

b)The formation of a non-perturbative condensate is also supported by
 observation of the phenomenon of coarsening, i.e.  the phenomenon of   
amplification of the zero mode as time increases. The effect of 
coarsening as well as the formation of a nonperturbative condensate is 
very similar to earlier discussions in ref.\cite{Boyanovsky}.

c)The formation of a non-perturbative condensate is also supported by 
the test of  volume -independence  of our results.  Namely, our 
quasi zero mode approaches the same value irrespective of the total
 spatial volume indicating that this really is a nonperturbative  
condensate that we are evolving to.  If it were not we would 
expect the value of the coefficient to decrease  when volume 
of the system increases.  

\section{Signatures and observables of  the produced 
$|\theta\ra^{ind}$ vacua} 
Having discussed the properties of  the $\theta$ vacuum
state itself, pseudo-Goldstone bosons as its lightest
excitations, and having assumed that this
unusual $|\theta\ra^{ind}$ ground 
state of matter is produced and will live a sufficiently
long time $\tau\sim(10-100 F)$,
we are now in position to answer   the question
formulated in the Introduction:  ``What  should  we
measure to  see all these
interesting phenomena?''.
In what follows we suggest
(based on our previous discussions
regarding the mass-shifts and other
properties of  the Goldstone modes)
a few simple observables 
which (hopefully) can be measured
and which uniquely signal that
the induced $|\theta\ra^{ind}$
vacuum state  indeed has been produced.
 
Before we begin, a few general remarks are in order.
 First of all, the produced 
$|\theta\ra^{ind}$ is a relatively short-lived  state
 with respect to electromagnetic and weak  interactions 
and  a relatively long-lived  state 
with respect to strong  interactions with
$\tau\sim (20MeV)^{-1}-
(2MeV)^{-1}$.
After the time $\tau$, the shell separating
$|\theta\ra^{ind}$ world from our world with $\theta=0$, breaks 
down\footnote{For numerical estimates presented below we use 
$\tau\simeq (5 MeV)^{-1}$.}.
All hadrons which have been formed in the $|\theta\ra^{ind}$ 
background will suddenly find themselves  in the new vacuum state
 (with $\theta=0$).   They have no choice but to   transform to 
the asymptotic states of this new (for them) $\theta=0$ world. 
The  dynamics of this strong transformation is quite complicated 
and beyond the scope of  the present work.  However, it is 
absolutely clear that the only signal which might be  relevant
 for  the study of $|\theta\ra^{ind}$-physics is a signature
 which  originates during  the time $\tau$ while $|\theta\ra^{ind}$ exists.
If something happens after that time, it can not be relevant for a study
 of the $|\theta\ra^{ind}$ vacua.
Our second remark is that a signal must  safely penetrate  through
the debris of the hot shell to deliver the information
about the induced  $|\theta\ra^{ind}$ state and not about something 
else. Essentially this constraint implies that the  decay products
originated during the time $\tau$ 
are better to be photons, electrons or  muons, rather than 
the strongly interacting hadrons. In this case
the effect  related to 
the electromagnetic rather than  strong interactions would be 
suppressed by a factor $\tau/ \tau_{em}\sim \alpha \ll 1$, where $ \tau_{em}$ 
is a typical  time scale for an electromagnetic process.
Based on these general remarks
and  previous discussions (Section 2)
regarding the mass-shifts (\ref{mass}) and parity-mixing  of the 
 hadrons in the $\theta$ background, we list  below and comment on some 
signatures which  (hopefully) can be measured and would signal  
 the  $|\theta\ra^{ind}$  vacua creation in Heavy Ion Collisions.

\subsection{ Signatures from the hadron decays  
in the   $|\theta\ra^{ind}$  background } 

{\bf a.} We start  from the analysis of 
$\eta\rightarrow\pi\pi$ as the most important 
signature of the produced $|\theta\ra^{ind}$ vacua.
As is known, this decay is strongly
suppressed in our world due to the 
opposite  CP parities of the initial ($\eta$) 
and  final states ($2\pi$).
In the previous section
we have  estimated the width (\ref{rate}) for the decay
 $\Gamma(\eta\rightarrow\pi\pi) 
\sim 0.5~MeV (\sin\frac{\theta}{2})^2$
  it turns out 
to be much larger than 
the full width $\Gamma(\eta)\simeq 1.18\pm 0.11 KeV$ of the  
$\eta$ meson in our world with $\theta=0$.
One should expect that  a noticeable fraction,
$\sim 0.5~MeV (\sin\frac{\theta}{2})^2\times\tau\sim 0.1$,  of 
all produced  $\eta$ mesons in the $\theta^{ind}$ background 
would decay to $2\pi$ before the shell separating the
 $\theta^{ind}$ background and our world with $\theta=0$ breaks down.
If the transition of these $2 \pi$ mesons   (produced in
the $\theta^{ind}$ background )
to our $|\theta=0\ra$ background is an adiabatic process, then  one should expect   
these $2 \pi$ mesons become
the asymptotic states of our world with
the corresponding masses and quantum numbers $0^{-}$.  However, 
the effect that these $\pi$ mesons were produced from $\eta$ decay 
in the $|\theta\ra^{ind}$ background
should not be washed out and the corresponding resonance behavior 
in   $\pi \pi$ invariant mass  should be seen in the spectrum. We should remark at this 
point that a signal  is expected only for low momentum pions, because
 $|\theta\ra^{ind}$ formation is caused by amplification of low 
momentum modes.
As well, as we mentioned in the previous section,
$|\theta\ra^{ind}$ will be different for  each given event, 
such that the corresponding
average over the large number of events is zero.
However, a measurement on an event by event
basis should produce a non-zero result. We should also remark here that mass
 shifts for Goldstone bosons in  $|\theta\ra$ background  and $|-\theta\ra$ 
background are the same. In both cases  masses are lower than in our vacuum 
state $|\theta=0\ra$ (see (\ref{mass})).
Our last remark:
the fact that hadron  properties are different in an 
unusual environment (when temperature $T$
and/or chemical potential $\mu$ are not zero)
is not a new idea  (see e.g.\cite{hadrons}).
However, $|\theta\ra^{ind}$ vacua is a quite special environment: in 
this background the hadrons do not posses the definite quantum numbers, 
like $P$ parity.  Therefore, their properties could be drastically 
different  from the ones which we know from the Particle Data Booklet.

{\bf b.} One could repeat the same comments regarding  
$\eta'\rightarrow\pi\pi$  decay as a very profound signature of the 
produced $|\theta\ra^{ind}$ vacua.
As is known, this decay is also strongly
suppressed in our world due to the 
opposite  CP parities of the initial ($\eta'$) 
and  final states ($2\pi$). Besides that, 
the full width of $\eta'$ is quite small 
$\Gamma(\eta')\simeq 0.2 MeV$ in our world with $\theta=0$.
In this case the effect is even more profound than in
the previously discussed  $\eta\rightarrow\pi\pi$ decay
because a larger  fraction of  all produced  $\eta'$ mesons 
in the $\theta^{ind}$ background
would decay to $2\pi$
before the shell separating the $\theta^{ind}$ 
background from our world with
 $\theta=0$ breaks down.
Indeed, from  eq.(\ref{eta7})we estimate this 
fraction on the level $\sim 2~MeV 
(\sin\frac{\theta}{2})^2\cdot\tau\sim 1/3$.

{\bf c.}
As we discussed above,
the electromagnetic, rather than the strong processes
could provide  a much  better signature for studying
the $|\theta\ra^{ind}$ induced vacua.
Of course, the relevant effects 
would be  suppressed by a factor $\tau/ \tau_{em}\ll 1$, 
where  $ \tau_{em}$ is a typical  time scale for 
an electromagnetic process, but, the signal 
is expected to be sufficiently clean due to the
free penetrating of leptons and photons through
the hot debris of the hadrons. In particular, 
the  mass of $\eta'$ meson in $|\theta\ra^{ind}$ 
induced background (\ref{mass}) can be independently
 measured from $\eta'\rightarrow 
\rho(\omega)\gamma$ decay  along with  $\eta'\rightarrow 2\pi$ 
decay as  described above.
The corresponding rate is expected to be 
on the same level as  in our world with $\theta=0$,
  i.e.  $\Gamma(\eta'\rightarrow \rho(\omega)\gamma )
\sim 0.06 MeV $. Therefore, a  quite noticeable portion
of the produced $\eta' s$ will decay to $\rho(\omega)\gamma $.
As usual, we only take into account  decays which were originated
while the $|\theta\ra^{ind}$   vacua existed, therefore the
effect is of the order $\sim \tau\cdot 0.06 MeV\sim 10^{-2} $.
One could discuss similar  decays such as 
$\omega\rightarrow\pi^0(\eta)\gamma$, 
$\phi\rightarrow\pi^0(\eta)\gamma$, $\eta'(\eta , \pi^0)
\rightarrow 2\gamma  $,   etc. with the following general conclusion:
the hadron mass shifts   and the lack of definite
$P, CP$ quantum numbers would lead, in general,
to some deviation from the standard pattern.
In particular,  the positions of the resonances as well as 
the  polarization properties of photons emitted from 
these hadrons, are expected to be  quite different from
what  one could anticipate normally in our world 
with $\theta=0$.
 
\subsection{Signatures from  the $|\theta\ra^{ind}$ decay  }
We have been discussing so far   
 the signatures originating  from   the hadrons
 ( which are excitations in the $|\theta\ra^{ind}$ background).
Now we want to discuss a signature which may occur
 when the ground state $|\theta\ra^{ind}$   decays by itself.
We should pause here to recall some relevant properties of the
$|\theta\ra^{ind}$ state.

As   explained in the previous sections
we expect that, in general, an    induced $|\theta\ra^{ind}$-
vacuum state
would be created in   heavy ion collisions,
similar to the creation of the disoriented chiral condensate  
with an arbitrary isospin direction.
It should be a large 
domain ($L\simeq 10-100 F$) with a wrong
$\theta^{ind}\neq 0$ orientation
which will {\bf mimic} the physics of the world with     fundamental 
$\theta^{fund}\neq 0$. It is expected that this 
macroscopically large     domain 
would be quite stable (with respect to  the strong  interactions)
 with $\tau\sim (20MeV)^{-1}-
(2MeV)^{-1}$. This state
 can be understood as the  appearance
of a macroscopically large  ($10-100 F$)  domain  where 
the    flavor- singlet phase of the chiral condensate is strongly correlated.
This singlet phase is identified  with $\theta^{ind}$ phase. Indeed, 
  this  identification is  a direct consequence of the transformation properties of
the fundamental QCD lagrangian under $U(1)_A$ rotations
when the chiral singlet phase can be 
rotated away at  the cost of the appearance of the induced
$\theta^{ind}$ term. A different name for the same object
would be a `` zero (spatially - independent) mode  of the $\eta'$ field"
which exactly corresponds to the
  flavor-singlet spatial-independent  part of the chiral condensate phase.
However, we prefer to use the term $|\theta\ra^{ind}$
because  the symbol $\eta'$ is usually associated with the $\eta'$ meson 
(excitation) and not with a classical constant field (condensate) 
in a large domain\footnote{We would like to make a  comment
 regarding this terminology. In Ref.\cite{TDLee} T.D.Lee considered 
the possibility of T violation in strong interactions due to the
$\la\eta'\ra$ condensate. It is equivalent, in a modern context, 
to the induced $\theta^{ind}\neq 0$ angle in the large domain
as explained above.
For example, his $\la\eta'\ra$   condensate induces a nonzero 
electric dipole moment of the neutron, see also comments by
 D. Kharzeev, R.D.Pisarski, and M.Tytgat in  Physics Today\cite{TDLee}.}.   

Having made these short remarks regarding the 
$|\theta\ra^{ind}$ state, we are in position to answer
on the question: `` What happens to the induced  
$|\theta\ra^{ind}$ state when  it  blows apart?''.
 The obvious answer is that  it will mainly decay to
the neutral pseudo-Goldstone bosons   with low
momentum modes  $|\vec{k}|\sim L^{-1}$.
 A detailed analysis
of the spectrum and other properties of this decay
will be presented elsewhere\cite{bfz1}, but here
I want to make some estimates based on the simple
energetic considerations. As we discussed earlier (\ref{energy}),
the vacuum energy density in the $\theta$ state is greater than in the
$\theta=0$ vacuum state 
by the amount: $$\Delta E=
E_{vac}(\theta)- E_{vac}(\theta=0)=2m|\la \bar{q} q \ra |
(1-|\cos\frac{\theta}{2}|) .$$
When $|\theta\ra^{ind}$ state blows apart, the  energy
associated with this background will be released as 
the Goldstone bosons carrying the total energy:
\beq
\label{energy1}
\Delta \epsilon_{\theta}\simeq 2m_q|\la \bar{q} q \ra |
(1-|\cos\frac{\theta}{2}|)V\simeq 20 
\cdot\left(\frac{V}{(10 F)^3}\right)GeV ,
\eeq
where $V=L^3$ is $3d$ volume of the $|\theta\ra^{ind}$ background
measured in $Fermi$.
Therefore, given the $cm$ energy $\sqrt{s}=40 TeV$, only
a small fraction: 
$$\rho\sim \frac{\Delta \epsilon_{\theta}}{\sqrt{s}}\sim \frac{ 20 
 (\frac{V}{(10 F)^3}) GeV}{40 TeV}\sim 10^{-3}(\frac{V}{(10 F)^3}) $$
of the total collision energy will be released through
the decay   of the induced $|\theta\ra^{ind}$ state.
Two of the most profound features of the Goldstone bosons produced 
in this decay are the following:  
1. Due to the fact that the 
 $|\theta\ra^{ind}$ formation is caused by amplification of low 
momentum modes, we expect the spectrum of 
the Goldstone bosons from this decay to
be  strongly enhanced at low $|\vec{k}|\sim L^{-1}$; 2.
Due to the nonzero value for the topological density
$\langle \theta | \frac{\alpha_{s}}{8 \pi } G \tilde{G}
| \theta \rangle =-\frac{\partial V_{vac}(\theta)}{\partial\theta}=
-m_q|\la 0| \bar{q} q |0 \ra |\sin\frac{\theta}{2}$, see
(\ref{energy}), one could expect that some $P, CP$ correlations
would appear in this background, as discussed in \cite{Pisarski}.
However, as we argued earlier this signal may be suppressed
 by the final state interactions. 

Therefore  we suggest  a somewhat  different signature to observe 
the formation of the induced
 $|\theta\ra^{ind}$ domain. As   mentioned above, 
an appropriate  signal have to be of the electromagnetic origin to be able
to deliver  the information on the local conditions
at its emission site. We argue below that such a signal would be
the excess of photons and/or leptons  which will be produced
   through the decay   of the induced $|\theta\ra^{ind}$ state
with a spectrum   which is strongly peacked at 
very low $|\vec{k}|\sim L^{-1}$ and which  demonstrates some
unusual polarization properties.
   The electromagnetic energy 
which we are about to discuss
 originates from two different sources: 1. the {\it direct} photons/leptons
which are produced from  the   electromagnetic  vacuum energy 
stored in the domain$V$;  2. the {\it indirect } photons/leptons
which are produced from  the neutral  pseudo -Goldstone
bosons $\pi^0\rightarrow 2\gamma , ~
\eta\rightarrow 2\gamma ,~
\eta'\rightarrow 2\gamma ,~\eta\rightarrow 3\pi^0
\rightarrow\gamma's $ etc. These neutral pseudo-Goldstone bosons
are essentially the decay products  of the $|\theta^{ind}\ra$ state
constructed from  diagonal $\phi_i$ fields (\ref{13}).

First of all, let us discuss the direct photon/lepton production.
 We want to argue
that there is an {\it electromagnetic}  vacuum energy 
stored in the region where the induced 
nonperturbative vacuum condensates  
have nonzero values (\ref{energy},\ref{chiral}).
The first hint (that the expectation value for the 
electromagnetic field might not be zero ) comes from analysis
of the effective lagrangian (\ref{13})  including the  term
$\Delta V_{em}$ related to the {\it electromagnetic} anomaly, 
\beq
\label{em}
\Delta V_{em}=-\frac{\alpha}{4\pi}N_c \left(\sum_i\phi_i Q_i^2\right)
 F_{\mu\nu}\tilde{F_{\mu\nu}},
\eeq
where $\alpha=1/137$ is the fine structure constant, $F_{\mu\nu}$
is electromagnetic field, $Q_i$ are the quark charges.
Phases, $\phi_i$, in this formula 
are defined in the same way as before (\ref{13}).
One usually  uses the expression (\ref{em}) for
 the  description of the  $\pi\rightarrow 2\gamma , ~
\eta\rightarrow 2\gamma ,~
\eta'\rightarrow 2\gamma $ decays. In this case
the $\phi_i$ phases are nothing but the standard
Goldstone bosons (with appropariate normalization
$\phi\rightarrow\frac{\pi}{f_{\pi}}$). However, 
the effective lagrangian (\ref{13}) together with
the additional term (\ref{em}) has a perfectly  physical meaning
even when  the $\phi_i$ fields describe 
the vacuum state itself, not necessarilly    
Goldstone excitations ($\pi$ mesons). This  was exactly the point
of the original papers \cite{WV},\cite{Wit2} and our discussion in section (2.1). 
In this case, a nonzero $\la \phi_i \ra$ 
in a macroscopically large (but finite) domain might
imply   a nonzero value for the correlation
$\la \frac{\alpha}{4\pi}N_c F_{\mu\nu}\tilde{F_{\mu\nu}}\ra$
over the same region. In other  words, due to the electromagnetic
 interaction $ F_{\mu\nu}$ with the phase of the quark condensate $\la \phi\ra $, 
the electromagnetic field also becomes correlated over the same region.

Another, more direct  hint which also suggests that the electromagnetic field
has a large scale correlation in the presence of the quark condensate, 
is based on the observation that the operators $
\frac{\alpha N_c}{4\pi} F_{\mu\nu} {F_{\mu\nu}}$ and 
$m\bar{\Psi}\Psi $ do mix. 
This mixing is absolutely irrelevant for QCD itself,  where
much more important  gluon operator such as $G_{\mu\nu}^2$ exists.
Nevertheless, such  a $F^2\leftrightarrow m\bar{\Psi}\Psi $
  mixing implies that in QCD
where the quark fields condense,  the electromagnetic field
(due to the interaction with these condensed quarks $m\bar{\Psi}\Psi $)
  also becomes a
  correlated field in the nonperturbative QCD vacuum background. 
This electromagnetic vacuum energy together
with the standard (infinite) perturbative vacuum energy 
(due to the zero vacuum 
fluctuations ) merely causes an unobservable shift of the vacuum energy.
This effect can be safely  ignored in all cases except the one 
where  a macroscopically large domain is produced  and has
  ground state    properties  different from our vacuum.
In this case, the electromagnetic vacuum energy can be released
through the production of  photons when the induced
$|\theta^{ind}\ra$ states falls apart.

The simplest way estimate  the scale of this effect is to
estimate the difference between $\la F_{\mu\nu}F_{\mu\nu}\ra $ 
in our world with $\theta=0$ and with $\theta\neq 0$.
As was explained above, due to the mixing 
of the operators $ F_{\mu\nu}F_{\mu\nu} $
and 
$m\bar{\Psi}\Psi $, the problem is reduced
to the analysis of the chiral condensate as  a function of 
$\theta$, which was discussed earlier, see eq.(\ref{chiral}).
Therefore, for approximate solution (\ref{vacuum})
we arrive at the following estimate:
\beq
\label{em1}
 |\la \theta| \frac{1}{4}F_{\mu\nu}F_{\mu\nu}|\theta\ra |
- |\la 0| \frac{1}{4}F_{\mu\nu}F_{\mu\nu}|0\ra |
\sim |\frac{3\alpha}{2\pi}(Q_u^2+Q_d^2)\ln(\mu L)
   \left(1-\cos(\frac{\theta}{2})\right)
\la m_q\bar{q}q\ra |  ,
\eeq
where $\mu$ is the normalization point, and $L\sim 10 F$
is a typical dimensional parameter of the system.
The sign in the formula is easy to understand: the 
 electromagnetic vacuum energy for the $|\theta\neq 0\ra$ ground state
is higher than for the $|\theta=0\ra$ state  because it
   is proportional to the vacuum contribution of 
the   chiral condensate $ \la m_q\bar{q}q\ra $,
which has  the lowest value for $\theta=0$ state, see (\ref{energy}).

As we discussed earlier,
when the $|\theta\ra^{ind}$ state blows apart, the  energy
associated with this background will be released mainly as 
the neutral pseudo-Goldstone bosons carrying the total energy 
given by eq.(\ref{energy1}) such that only a small fraction, 
 $\rho \sim 10^{-3}$,
of the total collision energy will be released through
the decay   of the induced $|\theta\ra^{ind}$ state.
Our estimate (\ref{em1}) essentially says that only a small fraction 
of this energy will be released as electromagnetic energy 
through production of {\it direct } photons and leptons.
\beq
\label{em2}
\rho_{em}\sim |\frac{3\alpha}{2\pi}(Q_u^2+Q_d^2)\ln(\mu L)|\rho \sim 10^{-6}.
\eeq
One should  say  the formula (\ref{em2})  is
probably an overestimation 
of the effect because we assumed  in the derivation that   along with the formation of the induced
$|\theta^{ind}\ra $ state, the 
electromagnetic fluctuations   also
settle down as prescribed by  eq.(\ref{em1}) during 
the same,  relatively short period of time $\tau\sim 10 F$.
This  may or may not be the case.
A further analysis is needed to study a time-scale required to 
form the electromagnetic correlation (\ref{em1}) after the chiral quark
condensate is formed. In any event, this effect is relatively small
and will not play an important role in the  discussion which follows.

Now we want to discuss the second (much more important) source 
of the electromagnetic energy which will be released
when $|\theta^{ind}\ra$ state
blows apart. This source  is related to the total energy of the background
(\ref{energy1}), rather than a  small electromagnetic fraction (\ref{em1}) of it.
Indeed, as we mentioned earlier, the  $|\theta^{ind}\ra$ domain itself
will eventually decay to the neutral pseudo-Goldstone bosons
$\pi^0, \eta, \eta'$, the fields which essentially form
the  $|\theta^{ind}\ra$ domain, see (\ref{13}).
It is important to note  that these neutral $\pi^0, \eta, \eta'$
bosons  will mainly    decay to photons/dileptons
such that large fraction of the   
background  energy   (\ref{energy1}) eventually 
will be released as the electromagnetic  energy.
Indeed, almost all produced $\pi^0 $ will decay to photons/dileptons;
more than 70   percent of $\eta $-mesons  will do the same through the decays $\eta\rightarrow
2\gamma, \eta\rightarrow 3\pi^0$. A noticeable fraction of $\eta^{\prime} $ mesons  will also transform 
to  the photons/dileptons through  the decays $\eta^{\prime} \rightarrow 2, \gamma , ~
\eta^{\prime} \rightarrow  \pi^0\pi^0\eta$. Therefore, we expect that  a  considerable part
of the vacuum energy (\ref{energy1}) stored in the $|\theta\ra^{ind}$ domain
will  be eventually released as {\ it indirect}  photons/dileptons.  

A  detailed discussion   of the spectrum of the produced particles will be presented
elsewhere\cite{bfz1}, but now let me mention that
 the most profound feature of the  
photons/dileptons   produced 
in the decay  of $|\theta\ra^{ind}$
domain,    is quite similar to what we discussed above  for the pseudo-Goldstone boson
 production themselves.  Namely,  due to the fact that the 
 $|\theta\ra^{ind}$ formation is caused by amplification of low 
momentum modes, we expect the spectrum of 
the produced particles  to
be  strongly enhanced at low $|\vec{k}|\sim L^{-1}$. We also expect that the  produced 
particles  may have some  unusual polarization (P, CP-odd) 
 properties due to the nonzero value of the condensates
$ \la G_{\mu\nu}\tilde{G_{\mu\nu}}\ra $ and $\la \bar{\Psi}i\gamma_5\Psi\ra $
inside   the domain $V$. In particular, one could expect that
the number of low-energy ($k\sim L^{-1}$) right- and left- handed photons will be different.

To conclude this section regarding the electromagnetic energy stored in the nontrivial
background, we would like to make the following remark. 
  The  electromagnetic
 $F_{\mu\nu}^2$  is  always correlated with the quark condensate. Therefore,
the corresponding  energy may be released
  when  the quark condensate 
in a large domain (due to   unusual conditions)  is different from  its vacuum value. The 
$|\theta^{ind}\ra $ state is only one example where this  may happen.
Another example would be an unusual environment with nonzero
temperature and chemical potential $\mu$ where, as is  well known, 
the chiral condensate may have a different magnitude. In this case 
the stored electromagnetic
energy will also be released through the direct photon/lepton production
as discussed above.
The only difference in the estimate (\ref{em1})
would be a replacement
\beq
\label{em3}
\left(1-\cos(\frac{\theta}{2})\right)
|\la m_q\bar{q}q\ra | \Longrightarrow
\left[ |\la m_q\bar{q}q\ra_{(\mu =T=0)}|-|\la m_q\bar{q}q\ra_{ (\mu \neq T\neq 0)}| \right],
\eeq
at least for sufficiently small $T$ and $\mu$.

Let me conclude this work with the following speculative conjecture.
   The longstanding problem
regarding the strong enhancement for invariant di-lepton masses below
the $\rho$ mass  reported by various collaborations at CERN\cite{dileptons}
may be intimately related to the low-energy  indirect photon/dilepton  production 
described  above. Indeed, as we argued above, a noticeable fraction of the total
energy  (\ref{energy1}) will be eventually released as    electromagnetic energy.
The main feature of these photons/dileptons is a strong enhancement
of the low-energy modes because the corresponding spectrum of the neutral pseudo-Goldstone
bosons is peaked  at low $k\sim L^{-1}$.  Detailed calculations\cite{bfz1} are needed 
before this speculative conjecture can get  further support.
At least, the standard theoretical analysis (see e.g. a nice summary 
  in ref. \cite{dileptons1} and references therein for details)  fails to reproduce
the enhancement in the data. 
Let me conclude on this optimistic note/speculation.
\section{Acknowledgments}
 I wish to thank Dima  Kharzeev, Larry McLerran,  Robert Pisarski  and Edward Shuryak
for valuable comments. I am also grateful to 
Adrian Melissinos for useful discussions
regarding  the  feasibility of  the experiment.
 I also would like to thank the members of   BNL's
 theory group for their interest in  this work.
The research was supported in part by Canadian NSERC.

\end{document}